\documentclass[aps,prl,twocolumn,floatfix,superscriptaddress,showpacs]{revtex4}
\usepackage{graphicx}
\usepackage{amssymb}
\usepackage{amsmath}

\newcommand{\pdag}{{\vphantom\dagger}}
\newcommand\s[1]{{\scriptscriptstyle #1}}
\newcommand\rs[1]{{\scriptscriptstyle \rm #1}}

\begin{document}

\title{Dynamical Unbinding Transition in a Periodically Driven Mott Insulator}

\author{Fabian Hassler}
\affiliation{Theoretische Physik, ETH Zurich, CH-8093
Zurich, Switzerland}
\affiliation{Instituut-Lorentz, Universiteit Leiden, P.O.\ Box 9506, 2300 RA
Leiden, The Netherlands}

\author{Andreas R\"uegg}
\affiliation{Theoretische Physik, ETH Zurich, CH-8093
Zurich, Switzerland}
\affiliation{Department of Physics, The University of Texas at Austin, Austin,
Texas 78712, USA}

\author{Manfred Sigrist}
\affiliation{Theoretische Physik, ETH Zurich, CH-8093
Zurich, Switzerland}

\author{Gianni Blatter}
\affiliation{Theoretische Physik, ETH Zurich, CH-8093
Zurich, Switzerland}
\date{\today}

\begin{abstract}
We study the double occupancy in a fermionic Mott insulator at half-filling
generated via a dynamical periodic modulation of the hopping amplitude.
Tuning the modulation amplitude, we describe a crossover in the nature of
doublon-holon excitations from a Fermi Golden Rule regime to damped Rabi
oscillations. The decay time of excited states diverges at a critical
modulation strength, signaling the transition to a dynamically bound
non-equilibrium state of doublon-holon pairs.  A setup using a fermionic
quantum gas should allow to study the critical exponents.
\end{abstract}


\maketitle

We study the response of interacting fermions on a lattice due to periodic
modulation of the hopping amplitude $w$ (shaking), combining strongly
correlated and non-equilibrium physics in a Mott insulator. Specifically, we
analyze a half-filled Hubbard model (one particle per site) in the
intermediate-temperature regime and determine the generation of double
occupancy due to the shaking of a single bond; the remaining bonds assume the
role of a reservoir. We map this problem onto a well-known impurity problem, a
single impurity coupled to a band of delocalized states
\cite{fano:61,anderson:61}. The pumped bond plays the role of the impurity,
the excited doublon-holon states which propagate through the lattice
constitute the band states, and the strength $\Omega$ of the time-dependent
drive represents the hybridization $V$; the strong interactions of the Mott
state at half-filling is mapped onto an effective renormalized density of
states of the reservoir.

Dynamical generation of double occupancy (DGDO) in a Mott insulator was
studied previously for a uniform modulation of all bonds in a one-dimensional
(1D) system using numerical methods \cite{kollath:06, massel:09} and in a 3D
system using perturbation theory in $\Omega$ \cite{huber:09, sensarma:09}.
This has led to a description of DGDO in terms of Fermi's Golden Rule (FGR)
providing the rate by which the double occupancy is changed. On the other
hand, it was noted that for intermediate modulation strength ($\Omega$ of the
order of the non-interacting band width) a crossover or transition from FGR to
coherent Rabi oscillations should be observed \cite{hassler:09}.

Here, we focus on the intermediate to large $\Omega$ regime requiring us to
include the drive in a non-perturbative way. We find that, besides the FGR
regime and the regime of damped Rabi oscillations studied before, an
additional regime of undamped Rabi oscillations appears at very strong drive:
if $\Omega$ exceeds a critical strength $\Omega_\mathrm{c}$, the dynamically
generated doublon-holon pair has no time to separate but is coherently pumped
back to the initial state with one particle per site. For very strong
coupling, such a non-equilibrium state does not carry delocalized excitations
and is therefore insulating. In quantum optics, this corresponds to a
saturated atomic transition \cite{tannoudji}. Reducing the driving power below
$\Omega_\mathrm{c}$, the system undergoes a transition where the size of the
bound doublon-holon pair diverges and the two excitations unbind.  Thus, the
transition from the undamped to damped Rabi oscillations is an unbinding
transition. The reverse transition with increasing drive induces binding by
pushing an isolated bound state out of a continuous spectrum. This situation
resembles that of repulsively-bound atom pairs in an optical lattice as
recently observed in cold atomic gases \cite{winkler:06, strohmaier:09}.  The
specific model presented here is motivated by a recent experiment using
modulation spectroscopy as a probe in a system of cold Fermi atoms subject to
an optical lattice \cite{jordens:08}.

We begin with a discussion of the $d$-dimensional model system, a single
impurity (c) at energy $\varepsilon_\mathrm{c}$ coupled to a band with
dispersion $\varepsilon_\mathbf{k}$ centered around $\omega=0$ with band
width $2W= 4dw$; in the Hamiltonian \cite{mahan:4}
\begin{equation}\label{eq:ham}
  H_\mathrm{c}= \varepsilon_\mathrm{c} |\mathrm{c}\rangle\langle\mathrm{c}| 
  + \sum_\mathbf{k}  \varepsilon_\mathbf{k} 
  |\mathbf{k} \rangle \langle \mathbf{k} |
  + V \bigl( |0\rangle\langle \mathrm{c} |
  + | \mathrm{c} \rangle \langle 0| \bigr),
\end{equation}
$|\mathrm{c}\rangle$ and $|\mathbf{k} \rangle$ are the impurity and ($N$)
continuum states, and $|0 \rangle = \sum_\mathbf{k} | \mathbf{k} \rangle
/\sqrt{N}$ is a localized (Wannier) state within the continuum which is
coupled to the impurity through the hybridization $V$.  As the Hamiltonian
(\ref{eq:ham}) describes a single particle problem, the Green's function
$G_\mathrm{c} (\omega) = \langle \mathrm{c} | (\omega - H_\mathrm{c} )^{-1} |
\mathrm{c} \rangle$ of the localized state can be obtained explicitly,
\begin{equation}\label{eq:green}
  G_\mathrm{c} (\omega)  = [\omega - \varepsilon_\mathrm{c} - V^2
  G_\mathrm{b} (\omega)]^{-1},
\end{equation}
with $G_\mathrm{b} (\omega) = \langle 0 | (\omega - H_\mathrm{b} )^{-1} |0
\rangle = N^{-1}\sum_\mathbf{k}(\omega - \varepsilon_\mathbf{k})^{-1}$ the
local Green's function of the band; here, $H_\text{b}$ denotes the Hamiltonian
of the band states [second term in (\ref{eq:ham})].  The coupling of the
impurity to the continuum shifts its energy to $\bar \varepsilon$ deriving
from ($\mathcal{P}$ denotes the principal value)
\begin{equation}\label{eq:shift}
  \bar \varepsilon =  \varepsilon_\mathrm{c} +  \frac{1}{N} 
  \sum_\mathbf{k}
  \mathcal{P}
  \frac{V^2}{\bar \varepsilon - \varepsilon_\mathbf{k}}
\end{equation}
and renormalizes its weight $Z$; if the energy of the new level $\bar
\varepsilon$ resides within the band continuum, it acquires a finite width
$\Gamma$.  We concentrate on the situation where the (unrenormalized) energy
$\varepsilon_\mathrm{c}$ is within the band, $|\varepsilon_\mathrm{c}|
\leq W$.  For small hybridization $V \ll w$, the Green's function is
well approximated by replacing the band's Green's function by its value
at $\omega=\varepsilon_\mathrm{c}$ and we obtain a local spectral function
\begin{equation}\label{eq:fgr}
  A(\omega)=-\frac{1}{\pi} \mathrm{Im}\, G_\mathrm{c}(\omega+i0^+) 
  \approx \frac{ V^2 \rho(\varepsilon_\mathrm{c})}{(\omega-
  \bar \varepsilon)^2 +\Gamma^2/4},
\end{equation}
i.e., the oscillator strength resides in a single level with energy
$\bar\varepsilon \approx \varepsilon_\mathrm{c}$ broadened by the FGR result
$\Gamma = 2 \pi V^2 \rho(\varepsilon_\mathrm{c})$, where $\rho(\omega) =
- \mathrm{Im}\,G_\mathrm{b} (\omega + i 0^+)/\pi$ is the local density of
states at the origin (we set $\hbar = 1$).  Equation~(\ref{eq:fgr}) leads
to an exponential decay of $G_\mathrm{c}(t)\propto \exp( - \Gamma t /2)$
in time.

Increasing the hybridization $V$ reveals corrections to the pure exponential
behavior.  A better approximation is obtained by retaining the
$\omega$-dependence in $\rho(\omega)$ in the numerator of (\ref{eq:fgr}) but
not in the denominator \cite{tannoudji:level}. This generalized FGR result
explains the rounding of $G_\mathrm{c}(t) \approx 1- V^2 t^2/2$ at short times
and the algebraic saturation $G_\mathrm{c}(t) \propto 1/t^{\mu+1}$ at large
times; here, the exponent $\mu$ describes the behavior of the density of
states near the band edge, cf.\ below. At strong coupling, $V \gg w$, the
Green's function can be approximated by $G_\text{b}(\omega) \approx
\omega^{-1}$ and $A(\omega)$ exhibits two $\delta$-functions at $\omega = \pm
V$, each with weight $Z= 1/2$; these are bound states which do not decay.  The
oscillations of the Green's function $G_\mathrm{c}(t > 0) = - i \cos(V t)$ can
be interpreted as Rabi oscillations between the states $|\mathrm{c} \rangle$
and $|0\rangle$; dropping the second term $\propto w$ in (\ref{eq:ham}) as
compared to the last $\propto V$, leaves us with an effective two-level
Hamiltonian, thus elucidating this result.

The crossover between these two regimes (the main target of this letter)
depends on the shape of the local density of states $\rho(\omega)$.  Coupling
a level to an unstructured continuum, the damping, although vanishing as $V
\to \infty$, remains finite for any finite $V$ \cite{tannoudji}. Coupling
the level to a finite-width continuum, the result depends on the shape
of $\rho(\omega) \sim (W \mp \omega)^\mu$ near the (upper/lower) band
edge \cite{tannoudji:level}.  In a non-interacting system with quadratic
dispersion the power $\mu$ is $(d-2)/2$.  For $\bar \varepsilon$ close
(but outside) the band, the sum in (\ref{eq:shift}) scales as $\sim V^2
(W \mp \bar\varepsilon)^\mu$. In 1D, $\mu=-1/2$, the term diverges as
$\bar\varepsilon \to \pm W$ approaches the band and a solution exists for
arbitrary small $V$, i.e., however small the hybridization, there exist
two localized eigenstates, one above and one below the band. In 2D, $\mu=0$
and the term scales as $\sim V^2 \log (W \mp \bar\varepsilon)$ leading to
the same conclusion with states exponentially close to the band.  In 3D
with $\mu=1/2$, a finite hybridization $V > V_\mathrm{c}$ is necessary
to satisfy (\ref{eq:shift}) with $\bar\varepsilon$ outside the band.
The critical hybridization $V_\mathrm{c}$ marks an unbinding transition:
for $V< V_\mathrm{c}$ there is no bound state in the system and $G_c(t
\to \infty) \to 0$, whereas for $V> V_\mathrm{c}$ part of the spectral
weight remains trapped in a bound state for arbitrarily long time and
$G_\mathrm{c}(t \to \infty)$ remains finite (but oscillatory).

Below, we are interested in the DGDO signal of interacting lattice fermions
periodically driven across one bond. We map this problem onto the above
impurity problem, with the interaction and many-body aspects manifesting
themselves in (\textit{i}) the bath's density of states, which is replaced by
a two-particle continuum characterized by an exponent $\mu_2 = 2 \mu_1 +1 =
d-1$ for a noninteracting particles and $\mu_\text{BR}=2$ for strongly
correlated particles; (\textit{ii}) the initial state of the bond, which can
be occupied by a singlet or a triplet; (\textit{iii}) multiple excitations,
which are created subsequently. We describe the system of interacting lattice
fermions at half-filling by the Hubbard Hamiltonian
\begin{equation}\label{eq:HHubbard}
  H_\mathrm{H} = 
  -w \sum_{\langle i,j\rangle, \sigma} 
  \bigl(c_{i\sigma}^\dag c_{j\sigma}^\pdag
  + c_{j\sigma}^\dag c_{i\sigma}^\pdag \bigr)
  + U \sum_{i} n_{i\uparrow}n_{i\downarrow}
\end{equation}
and account for the harmonic modulation (frequency $\nu$) of the hopping $w$
over the bond $\langle1,2\rangle$ via
\begin{equation}
  H_\mathrm{osc}(t)= \Omega \cos(\nu t) \sum_{\sigma}
  \bigl( c_{1\sigma}^\dag c_{2\sigma}^\pdag 
        + c_{2\sigma}^\dag c_{1\sigma}^\pdag \bigr).
   \label{eq:Hosc}
\end{equation}
Here, $c_{i\sigma}^{\dag}$ are fermion operators creating particles with spin
$\sigma$ in a Wannier state at site $i$ and the sum $\langle i,j\rangle$ runs
over nearest neighbors for all $N$ lattice sites. The hopping amplitude is
denoted by $w$ and $U$ is the on-site interaction. We restrict our analysis 
the strongly correlated intermediate-temperature ($T$) regime $w^2/U \ll w
\sim k_\rs{B} T \ll U$, such that the initial state is a Mott-like state with
one particle per site. The oscillatory part, Eq.~(\ref{eq:Hosc}), generates
double occupancy by coupling a spin-singlet on the bond $\langle 1,2\rangle$
to a doublon-holon pair.

The DGDO signal is largest in the strongly correlated regime at half-filling
\cite{huber:09}. In this case it is appropriate to truncate the full Hilbert
space and keep only the two lowest-energy sectors $\alpha$ and $\beta$ of
(\ref{eq:HHubbard}) at fixed particle number. The low-energy sector $\alpha$
(with width $w^2/U \ll k_{\rm\scriptscriptstyle B} T$) is centered around the
energy $E=0$ and the eigenstates are well approximated by product states
$|\{s\}\rangle_\alpha$ describing singly-occupied sites with spin
configurations $\{s\}$. The next-higher energy sector $\beta$ (Hubbard bands)
is centered around the energy $E=U$. The eigenstates in $\beta$ involve a
(delocalized) empty site (holon) and a (delocalized) doubly occupied site
(doublon) which form a two-particle continuum.  We denote states of the
doublon-holon continuum by $|i_\rs{h},j_\rs{d}; \{s\}'\rangle_\beta$, where
$i$ ($j$) denote empty (doubly occupied) sites and $\{s\}'$ is the spin
configuration of the remaining sites.

The time-dependent part $H_\mathrm{osc}(t)$ couples the $\alpha$-sector to
the $\beta$ states \cite{hassler:09}. We find the average number of doublons
$D(t)=P_{\alpha\beta}(t)$ from the probability $P_{\alpha\beta}(t)$ for
a transition between the $\alpha$- and $\beta$-sectors; the latter can be
obtained via $P_{\alpha\beta}(t)=1-P_{\alpha\alpha}(t)$ from the persistence
\begin{equation}\label{eq:paa}
  P_{\alpha\alpha}(t)= \frac{1}{N_\alpha}
  \sum_{\{s\}} \bigl| 
  {}_\alpha\langle\{s\}| \mathcal{T}\,
e^{-i\int_{-\infty}^{t} \!\! dt' \, H(t')}|\{s\}\rangle_\alpha \bigr|^2,
\end{equation}
i.e., the probability to stay in $\alpha$; here $N_\alpha= 2^N$ is the number
of states in the $\alpha$-sector, $N$ the number of lattice sites, and
$\mathcal{T}$ denotes the time-ordering operator. In order to remove the
time-dependence, we go to the interaction picture with respect to the
reference Hamiltonian $H_0=- \nu \sum_{\{s\}}|\{s\}\rangle_\alpha\,
{}_\alpha\langle\{s\}|$.  The new Hamiltonian then reads
$H'(t)=U^{\dag}(t)[H(t)- H_0]U(t)$ with $U(t)=\exp(-iH_0 t)$. As a
consequence, all energies in the $\alpha$-sector are shifted by $\nu$ with
respect to the $\beta$-sector, such that the states overlap in energy for $\nu
\approx U$. Furthermore, we dispose of the time-dynamics in $H_\mathrm{osc}$
via a rotating wave approximation (dropping terms oscillating with frequency
$2\nu$) and obtain the new driving term (cf.\ Eq.\ (\ref{eq:ham}))
\begin{equation}\label{eq:coupl}
 H'_\mathrm{osc} = \Omega \sum_{\{ s\}'} \bigl( 
 |S;\{s\}'\rangle_\alpha \, {}_\beta \langle D;\{s\}'|
 + \mathrm{H.c.}
 \bigr) + B
\end{equation}
with the relevant states in the $\alpha$-, $\beta$-sector
\begin{align*}
  |S;\{s\}'\rangle_\alpha &
  = \bigl( |1_\s{\uparrow} 2_\s{\downarrow}; \{s\}' \rangle_\alpha
  - |1_\s{\downarrow} 2_\s{\uparrow}; \{s\}'\rangle_\alpha \bigr)/\sqrt{2},\\
  |D;\{s\}'\rangle_\beta &
  = \bigl(|1_\rs{h} 2_\rs{d}; \{s\}' \rangle_\beta
  + |1_\rs{d} 2_\rs{h}; \{s\}' \rangle_\beta \bigr)/\sqrt{2}.
\end{align*}
The operator $B$ conserves the number of doubly occupied and empty sites and
involves processes where empty/doubly occupied sites on the bond $\langle 1,2
\rangle$ interchange with singly occupied sites.

Assuming that the initial state is $|S,\{s\}'\rangle_\alpha$ with $\{s\}'$ an
arbitrary spin configuration, we are left with a single level coupled to a
continuum via a hybridization energy $\Omega$, i.e., the problem discussed
above. The role of the local Green's function $G_\mathrm{c}$ is played by
$\mathcal{G}_{\alpha\alpha}(t) = -i \Theta(t) \,{}_\alpha\langle S,\{s\}'|
\,\mathcal{T} \exp \bigl[-i\int^{t} dt' \, H(t')\bigr] |S,\{s\}'
\rangle_\alpha$, whose Fourier transform is given by (cf.\ Eq.~\ref{eq:green})
\begin{equation}\label{eq:Gcal}
  \mathcal{G}_{\alpha\alpha}(\omega)
  = \int \! dt\, e^{i\omega t} \mathcal{G}_{\alpha\alpha} (t) =
  \frac{1}{\omega - \delta-\Omega^2G(\omega)},
\end{equation}
with $G(\omega)$ the Fourier transform of the (local) two particle Green's
function of the $\beta$-sector
\begin{equation}\label{eq:Gt}
  G(t) = \frac{-i \Theta(t)}{N_\beta} 
  \sum_{\{s\}'} {}_\beta\langle D; \{s\}'|\mathcal{T} 
  e^{-iH_\mathrm{H} t}|D; \{s\}'\rangle_\beta
\end{equation}
replacing the Green's function $G_\mathrm{b}$ of the band.  Here, $N_\beta$ is
the number of states in sector $\beta$ and the detuning $\delta=\nu-U$
replaces $\epsilon_\mathrm{c}$; below, we limit ourselves to the case
$\delta=0$ which describes the situation near $\nu \approx U$.

Once $G(\omega)$ is known, we can calculate the probability $P_{\alpha\beta}
(t)= 1- P_{\alpha\alpha}(t)$ for a transition between the two sectors using
Eq.~(\ref{eq:Gcal}) and
\begin{equation}\label{eq:paa2}
  P_{\alpha\alpha}(t) = \bigl[3 + 
  |\mathcal{G}_{\alpha\alpha}(t) |^2
  \bigr]/4;
\end{equation} 
the two terms refer to triplet- (probability $3/4$) and singlet-
($|S;\{s\}'\rangle_\alpha$) type initial states.  In calculating the
two-particle Green's function Eq.~(\ref{eq:Gt}), we make use of particle-hole
symmetry and treat the doublon and the holon as independent particles,
$G(t)\approx i g(t)^2$, except for the initial correlations between the
doublon at $1$ and the holon at $2$.  For the one-particle Green's function
$g(t)$ we make use of the retraceable path approximation due to Brinkman and
Rice (BR) \cite{brinkman:70}, with the additional prescription to exclude site
$2$ in the first hopping process as doublon-holon exchange is suppressed by
$w/U$. Every hop then is equivalent and the original BR result simplifies,
generating the single particle Green's function $g(\omega)=2\omega
\omega_0^{-2} [1-\sqrt{1 - (\omega_0/\omega)^2}]$, where $\omega_0=2w
\sqrt{2d-1}$. The two-particle Green's function is the convolution
$G(\omega)=i\int \!dz \,g(\omega-z)g(z)$ and we obtain
\begin{equation*}
  G(\omega)= \frac{4 x}{\omega_0} \biggl\{1+ \frac{4}{3\pi} \bigl[ (x^2 - 1)
  \mathrm{K}(x^{-1}) + (x^2+1) \mathrm{E}(x^{-1}) \bigr] \biggr\}
\end{equation*}
with the corresponding two-particle density of states
\begin{equation*}
  \rho(\omega)=\frac{16}{3\pi^2\omega_0}\biggl[ (x^2+1)
  \mathrm{E}\bigl(\sqrt{1-x^2}\bigr) \\
  -2x^2 \mathrm{K}\bigl(\sqrt{1-x^2}\bigr)\biggr]
\end{equation*}
where $x=\omega/2\omega_0$ and $\mathrm{K}(x)$ $[\mathrm{E}(x)]$ is the
complete elliptic integral of the first [second] kind \cite{abramoviwtz}. 

\begin{figure}
\centering
\includegraphics[width=0.95\linewidth]{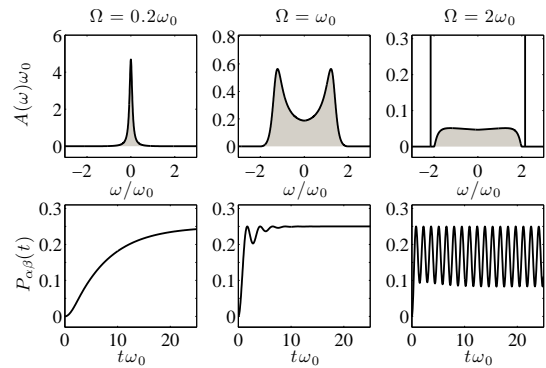}
\caption{%
(Top row) The local spectral function $A(\omega)$ of the $\alpha$-sector for
different values of the modulation strength $\Omega$. (Bottom row) The
probability for a doublon-holon pair $P_{\alpha\beta}(t)$ as a function of
time $t$. From left to right, the three different regimes are characterized by
the (generalized) Fermi's golden rule, damped, and undamped Rabi oscillations.
}
\label{fig:AandPab}
\end{figure}

\begin{figure}
\centering
\includegraphics[width=0.84\linewidth]{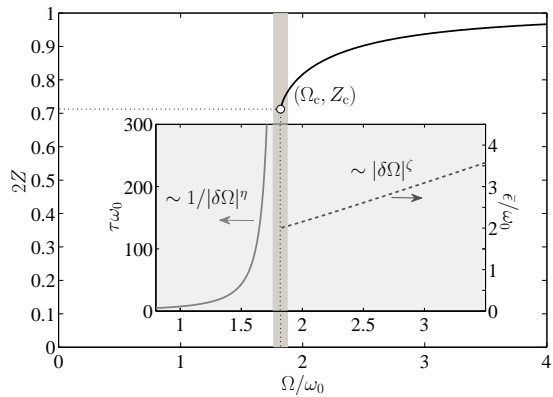}
\caption{%
Characteristics of the dynamical unbinding transition. The plot shows the
weight $Z$ of the coherent oscillations as a function of the driving strength
$\Omega > \Omega_\mathrm{c}$. At the critical driving strength
$\Omega_\mathrm{c}$ the weight $Z$ drops from the finite value $Z_\mathrm{c}$
to 0 indicating the unbinding transition. The inset shows the divergence of
the correlation time $\tau$ for $\Omega< \Omega_\mathrm{c}$ (with $\eta =
2$) as well as the shift of the persistent oscillation frequency
$\bar\varepsilon$ for $\Omega > \Omega_\mathrm{c}$ (with $\zeta =1$).
}
\label{fig:Zandtau}
\end{figure}

The weak to strong driving crossover is governed by the behavior of the
density of states near the band edge. Expanding $\rho(\omega)$ around the band
edge at $2\omega_0$, we find an exponent $\mu_\text{BR}=2$ and hence the
strongly correlated doublon-holon continuum features a transition from the FGR
regime to Rabi oscillations at a critical drive $\Omega_\mathrm{c}>0$. We
focus on the local spectral function $A(\omega)$, cf.~(\ref{eq:fgr}), from
which the propagator $\mathcal{G}_{\alpha\alpha} (t)=-i\int d\omega
e^{-i\omega t} A(\omega)$ and the (directly measurable) number
$P_{\alpha\beta}(t)$ of doublons are easily derived, cf.\ Fig.\
\ref{fig:AandPab}.
The limits of weak and strong driving lead to the similar results as described
in Refs.~\cite{sensarma:09,hassler:09} (see also the discussion
below~(\ref{eq:shift})).  Here, we concentrate on the regime near the
transition with $\Omega \approx \Omega_\mathrm{c}$.  This transition between
different non-equilibrium states is dynamically triggered and marks the
unbinding of the localized doublon-holon pairs. The critical modulation
strength $\Omega_\mathrm{c}$ is obtained by solving Eq.\ (\ref{eq:shift}),
$\bar\varepsilon=\Omega_\mathrm{c}^2 \,\mathrm{Re}\,G(\bar\varepsilon)$ near
the band edge $\bar\varepsilon \rightarrow 2 \omega_0$.  For $\omega\geq
2\omega_0$, we find $\Omega_\mathrm{c} = \omega_0/\sqrt{2-16/3\pi}\approx
1.82\,\omega_0$;  approaching $\Omega_\mathrm{c}$ from above, we find that
$(\bar\varepsilon-2\omega_0) \sim (\Omega-\Omega_\mathrm{c})^{\zeta}$ with the
exponent $\zeta$ quantifying the level repulsion; here, $\zeta=1$ and levels
do not repel. At criticality $\Omega_\mathrm{c}$, the oscillator strength in
the two $\delta$-functions is $2Z_\mathrm{c}=3\pi/2-4\approx 0.71$.  For
$\Omega < \Omega_\mathrm{c}$, we trace the poles in the analytic continuation
of Eq.\ (\ref{eq:shift}) to the lower complex half-plane and determine the
imaginary parts providing the inverse decay time of damped Rabi oscillations,
$\tau^{-1}=-2 \, \mathrm{Im}\,\bar \varepsilon$.  Approaching
$\Omega_\mathrm{c}$ from below leads to a diverging decay time $\tau\sim
|\Omega-\Omega_\mathrm{c}|^{-\eta}$ with $\eta=2$; an overview of these
characteristics is presented in Fig.~\ref{fig:Zandtau}.
The above incoherent hopping in a random spin background above is basically
independent of dimensionality (except for $\omega_0$); when hopping is in a
polarized background, the coherent propagation of doublons/holons is given by
the non-interacting band-dispersion and we expect a qualitatively different
behavior in different dimensions, e.g., $\Omega_\mathrm{c} =0$ in 1D where
$\mu_2=0$.

In restricting the many-body Hilbert space to the $\alpha$ and
$\beta$-sectors, we have neglected excitations which involve two or more
doublon-holon pairs. Such processes occur when the pair leaves the bond
$\langle 1,2\rangle$ within the time $1/\Omega$, which is the case for $w >
\Omega$; the bond then is ready for a new excitation.  When $\Omega \gg w$
(the Rabi oscillation regime), the doublon-holon pair remains localized
around the sites 1,2 and thus blocks further excitations.  In the opposite
(FGR) limit ($\Omega \ll w$), the assumption of a single excitation is not
valid and we need to include further excitations; within the FGR regime
these are incoherent and independent.  The probability that (at least)
a single doublon-holon pair is created is given by $P_{\alpha\beta} = [1 -
|\mathcal{G}_{\alpha\alpha}(t)|^2]/4 \equiv P_1$, cf.\ Eq.~(\ref{eq:paa2}).
Assuming independent processes, the probability $P_n(t)$ that $n$
(or more) doublon-holon pairs are created is given by $P_n(t)= P_{n-1}
[1-|\mathcal{G}_{\alpha\alpha}(t)|^2]/4$ and we find an average number
of doublons
\begin{equation}\label{eq:d}
   D(t) = \sum n p_n = \sum_n P_n = \frac{1 -
   |\mathcal{G}_{\alpha\alpha}(t)|^2}{3 + |\mathcal{G}_{\alpha\alpha}(t)|^2};
\end{equation}
here, $p_n$ denote the probability that $n$ doublon-holon pairs are created.
The previous relation $D(t)=P_{\alpha\beta}(t)=[1-|\mathcal{G}_{\alpha\alpha}
(t)|^2]/4$, Eq.~(\ref{eq:paa}), thus is only valid for $|\mathcal{G}_{\alpha
\alpha} (t)|^2 \approx 1$, i.e., at short times. Also, equation (\ref{eq:d})
describes the saturation at a value $D =1/3$ for long times.

The system discussed in the present letter can be realized, e.g., in a setup
with fermionic cold atoms subject to an optical lattice and interactions tuned
by a Feshbach resonance. The dynamical driving of a single bond $\langle 1,2
\rangle$ can be achieved by an additional laser field modulating the potential
well between the sites; note though, that we are interested in high-driving
powers where $\Omega > w$. As it is impossible to achieve negative tunneling
amplitudes, the bond $\langle 1,2 \rangle$ then needs to be driven around a
static value $w_{12} \gg w$ of the hopping amplitude.

We thank I.\ Carusotto and S.D.\ Huber for stimulating discussion and
acknowledge financial support from the Swiss National Foundation through
the NCCR MaNEP.


\begin{thebibliography}{10}

\bibitem{fano:61}
U.\ Fano,
 Phys.\ Rev.\ {\bf 124}, 1866 (1961).

\bibitem{anderson:61}
P.W.\ Anderson,
 Phys.\ Rev.\ {\bf 142}, 41 (1961).

\bibitem{kollath:06}
C.\ Kollath, A.\ Iucci, I.P.\ McCulloch, and T.\ Giamarchi,
 Phys.\ Rev.\ A {\bf 74}, 041604(R) (2006).

\bibitem{massel:09}
F.\ Massel, M.J.\ Leskinen, and P.\ Torma,
 Phys.\ Rev.\ Lett.\ {\bf 103}, 066404 (2009).

\bibitem{huber:09}
S.D.\ Huber and A.\ R\"uegg,
 Phys.\ Rev.\ Lett.\ {\bf 102}, 065301 (2009).

\bibitem{sensarma:09}
R.\ Sensarma, D.\ Pekker, M.D.\ Lukin, and E.\ Demler,
 Phys.\ Rev.\ Lett.\ {\bf 103}, 035303 (2009).

\bibitem{hassler:09}
F.\ Hassler and S.D.\ Huber,
 Phys.\ Rev.\ A {\bf 79}, 021607(R) (2009).

\bibitem{tannoudji}
C.\ Cohen-Tannoudji, J.\ Dupont-Roc, and G.\ Grynberg,
 {\em Atom-Photon Interactions}
 (Wiley, New York, 1992).

\bibitem{winkler:06}
K.\ Winkler {\em et~al.},
 Nature {\bf 441}, 853 (2006).

\bibitem{strohmaier:09}
N.\ Strohmaier {\em et~al.},
accepted in  Phys.\ Rev.\ Lett.\ (arXiv:0905.2963).

\bibitem{jordens:08}
R.\ J\"ordens {\em et~al.},
 Nature {\bf 455}, 204 (2008).

\bibitem{mahan:4}
G.D.\ Mahan,
 {\em Many-Particle Physics}, chapter 4.2,
 3rd~edition (Plenum, New York, 2000).

\bibitem{tannoudji:level}
Chapter $C_\mathrm{III}$ in  \cite{tannoudji}.

\bibitem{brinkman:70}
W.F.\ Brinkman and T.M.\ Rice,
 Phys.\ Rev.\ B {\bf 2}, 1324 (1970).

\bibitem{abramoviwtz}
M.\ Abramowitz and I.A.\ Stegun, eds.,
 {\em Handbook of Mathematical Functions},
 (Dover, New York, 1972).

\end{thebibliography}
\end{document}